\documentclass[a4paper,11pt]{article}
\usepackage{pos}
\usepackage{subcaption}
\usepackage{lineno}

\title{Measuring total neutrino cross section with IceCube at intermediate energies ($\sim$100 GeV to a few TeV)}
 \ShortTitle{IceCube: intermediate energy neutrino cross section}

\author{The IceCube Collaboration \\{\normalsize \normalfont(a complete list of authors can be found at the end of the proceedings)}}

\emailAdd{snowicki@icecube.wisc.edu}

\abstract{Whether studying neutrinos for their own sake or as a messenger particle, neutrino cross sections are critically important for numerous analyses. 
On the low energy side, measurements from accelerator experiments reach up to a few 100s of GeV. 
On the high energy side, neutrino-earth absorption measurements extend down to a few TeV. 
The intermediate energy range has yet to be measured experimentally. 
This work is made possible by the linear relationship between the event rate and cross section, and will utilize IceCube muon neutrino data collected between 2010 and 2018. 
An advanced energy reconstruction, tailored to the unique properties of the energy range and using the full description of photon propagation in ice, is applied to an event sample of neutrino-induced through-going muons to perform a forward folding analysis.\\

\vspace{4mm}
{\bfseries Corresponding authors:}
Sarah Nowicki$^{1*}$\\
{$^{1}$ \itshape Michigan State University}\\[4mm]
$^*$ Presenter

\FullConference{37$^{\rm{th}}$ International Cosmic Ray Conference (ICRC 2021)\\
		July 12th -- 23rd, 2021\\
		Online -- Berlin, Germany}

}

\begin{document}
\maketitle

\section{Motivation}\label{sec:motive}

Neutrino-nucleon cross sections are a topic of fundamental importance to many neutrino experiments. 
Accelerator-based neutrino experiments have made extensive measurements with increasing precision in the complex 
energy region up to 360 GeV \cite{Zyla:2020zbs}. 
More recently, higher energy measurements have become a reality with the availability of samples of 
astrophysical and atmospheric neutrinos from the IceCube Neutrino Observatory at TeV - PeV energies \cite{Aartsen:2017kpd, Robertson:2021icrc, Abbasi:2020luf, Xu:2017ufb, Bustamante:2017xuy}.  
However, no experimental measurements currently exist in the intermediate energy range of few hundred GeV to a few TeV. 
IceCube also has the capacity to measure high statistics samples of high quality events at these energies. 
Building on a pre-existing, highly developed event selection \cite{Stettner:2019tok} and tools allows the first measurement of the charged current (CC), total $\nu_{\mu}$ cross section in this critical regime. 

\section{Detector and Event Sample}\label{sec:sample}

The IceCube detector consists of a cubic-kilometer array of 5160 optical detectors embedded in the Antarctic ice near the South Pole Station \cite{Aartsen:2016nxy}. 
Charged particles, including those produced by neutrino interactions with the nuclei in the surrounding ice or nearby bedrock, may be detected via their emitted Cherenkov radiation. 
The three-dimensional arrangement of sensors, as well as charge and timing information of detected photons, allows us to reconstruct particle properties such as energy, direction and type. 

For this cross section analysis, a high statistics sample of well-understood muons derived from $\nu_{\mu}$ CC interactions with minimal background is desired. 
Such an event sample already exists for the purposes of studying the diffuse flux of astrophysical neutrinos \cite{Stettner:2019tok}. 
The event selection chooses 
through-going muon tracks originating outside of the detector. 
It selects events coming from the bedrock side of the detector, utilizing the earth as a shield as well as a Boosted Decision Tree to remove atmospheric muon backgrounds to achieve 99.7\% neutrino purity. 
This analysis will use data collected from May 2010 to December 2018. 
The diffuse astrophysical analysis also included data from May 2009 - May 2010, but due to a difference in calibration, it must be treated separately from the other years of data \cite{Stettner:2019tok}. 
Given that the year has a limited statistical contribution, it is excluded for simplicity. 


\section{Event Reconstruction}\label{sec:reco}

One downside to the existing event sample is that it uses an energy reconstruction geared to the high-energy end of the spectrum, the focus of the astrophysical neutrino flux analysis. 
For reference, this method uses the truncated mean of muon dE/dx as an energy estimator \cite{Abbasi:2012wht}. 
For the purposes of this GeV - TeV focused study, a new reconstruction was tailored specifically for the low-energy part of the event sample. 
We use a log-likelihood-based fitter as the starting point, initially developed to utilize pre-generated templates of charge prediction for unit events, which may be scaled or stacked to estimate particle energy \cite{Aartsen:2013vja}. 
A limitation of using these templates, or photon lookup tables, is that they are specific to the modelling of the glacial detector medium. 
The level of complexity captured in the lookup tables is limited by the requirements not only to generate and store the tables, but also to load them every time an event is reconstructed. 
IceCube ice models have progressed well beyond the level of detail captured in the most advanced tables and continue to be an active 
area of study \cite{RongenChirkin:2021icrc, Chirkin:2021icrc}. 
Instead, we generate expected charge for event hypotheses via OpenCL-based photon propagation simulation software. 
This grants us additional flexibility in both our choice of ice model and construction of hypothesis event. 
The method is referred to as DirectReco for its direct use of photon propagation in reconstruction. 

We can use that flexibility to tune our representation of our target events. 
Our goal is to reconstruct the energy of through-going muons, that is, muons entering the instrumented volume from outside of the detector. 
Our event hypothesis uses a series of point-like emissions to approximate the muon track. 
To decide how the muon energy should be distributed, we consider how the energy deposition of a muon changes with energy. 
For energies below $\sim$100 GeV, the muon is a minimum ionizing particle, depositing a constant amount of energy as it traverses the detector. 
At a few TeV and above, muon energy deposition becomes dominated by radiative losses caused by stochastic interactions. 
Accordingly, 
the point-like losses that represent the muon track are spaced equidistantly, each one has an equal fraction of the total track energy, and the likelihood is optimized with respect to that total track energy. 

\begin{figure} \centering
\includegraphics[width=0.6\textwidth]{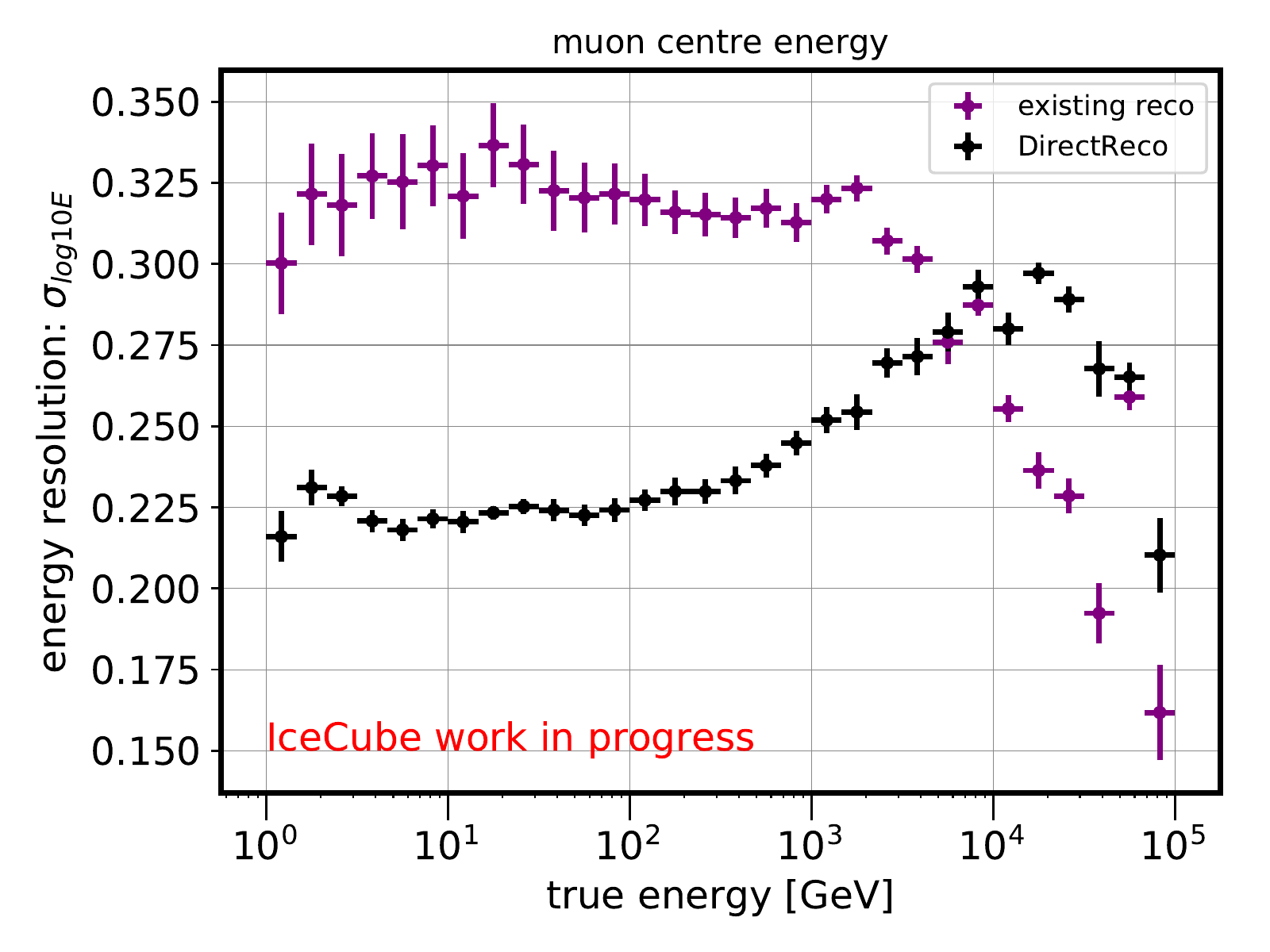}
\caption{Resolution plot comparing existing reconstruction\cite{Abbasi:2012wht} from diffuse analysis and DirectReco for a subset of final level Monte Carlo events. 
True energy here is the muon energy at its closest point to the centre of the detector. 
DirectReco shows a significant improvement in resolution at a few TeV and below, the target energy for this analysis.}
\label{fig:resolution}
\end{figure}

Comparing the performance of the two methods in Fig.~\ref{fig:resolution}, we see an improvement in resolution for DirectReco from the lowest energies up to a few TeV, where 
muon energy losses become stochastics dominated and hence the existing method begins to improve. 
This study was performed using Monte Carlo events processed through the analysis event selection. 
DirectReco's enhancement of energy resolution for the critical GeV - TeV energy window improves the precision of this cross section measurement.

This version of DirectReco for through-going muons is not optimized for energies above a few TeV, given that 
the hypothesis is simply not a good physical description at these higher energies. 
Additionally, reconstruction time and memory required increase with event energy in general. 
To reduce those constraints, we introduce a condition which aims to process only low-energy ($<$5 TeV) events while retaining the existing reconstruction for high-energy ($>$5 TeV) events. 
We choose the observable of deposited charge (100 photo-electrons) as an energy proxy to make this determination. 

\section{Analysis}\label{sec:analysis} 

TeV to PeV neutrino-nucleon cross section analyses utilize Earth absorption to infer the cross section for a given model of Earth's interior.  
However, as neutrino energy decreases, the effect lessens until the Earth becomes transparent to neutrinos around 1 TeV. 
This analysis will instead rely on the linear relationship between event rate and cross section, making a measurement via normalization of predicted neutrino flux.

We build on the astrophysical flux analysis associated with the event sample \cite{Stettner:2019tok} 
as our foundation. 
Sources considered 
include conventional and prompt atmospheric neutrino fluxes from cosmic ray interactions and an isotropic flux of astrophysical neutrinos, as well as a sub-dominant contribution of mis-reconstructed atmospheric muons. 
Atmospheric neutrino fluxes are modelled using MCEq software \cite{Fedynitch:2015zma}, with the Gaisser-Hillas (H4a) cosmic ray model \cite{Gaisser:2012zz} and SYBILL2.3c hadronic interaction model \cite{Fedynitch:2018cbl} as the baseline.
Other models are considered via nuisance parameters. 
Monte Carlo simulations were generated for the various sources included and will be compared  to experimental data in reconstructed space via a Poisson likelihood. 
The events are binned in estimated muon energy and cosine of the reconstructed muon zenith angle.

We will measure the cross section as a multiplicative factor on the Standard Model theoretical model for these energies, CSMS \cite{CooperSarkar:2011pa}, used in our Monte Carlo simulations. 
Two physics parameters are fitted; one governs events at 100 - 350 GeV and the other at 350 GeV - 5 TeV. 
Since IceCube has no mechanism to distinguish between neutrinos and antineutrinos, the multiplicative parameters apply to their sum. 
However, the cross sections are quite different here, 
e.g. at 500 GeV, $\sigma_{\nu}/\sigma_{\bar{\nu}} \sim 1.78$. 
It is understood that this may have an important effect in the energy region under study and is being evaluated.

An 
important element is the treatment of the atmospheric neutrino flux that dominates 
our energy range. 
To describe the uncertainties from the cosmic ray modelling, two nuisance parameters are implemented. 
One modifies the cosmic ray spectral index and the other interpolates between two disparate cosmic ray models, H4a \cite{Gaisser:2012zz} and GST4 \cite{Gaisser:2013bla}, to represent shape differences.  
To describe the uncertainties from the hadronic interaction modelling, four parameters from the Barr parameterization \cite{Barr:2006it} are implemented, 
specifically, $H^{\pm}$, which modifies pion production, and $W^{\pm}$, $Y^{\pm}$, $Z^{\pm}$, which modify kaon production. 
The Barr parameters have Gaussian priors with widths determined by the uncertainties derived in the paper. 
Varying all six parameters within 1 sigma gives the fit sufficient flexibility to describe many 
different atmospheric flux models \cite{Stettner:2021xxt}. 

The conventional atmospheric neutrino flux normalization is highly correlated to our cross section parameters.
To partially break the correlation, we use existing accelerator-based cross section measurements \cite{Zyla:2020zbs} for energies 100 - 360 GeV to construct a prior. 
We use the data points, given as $\sigma/E_{\nu}$, to calculate an error-weighted mean and standard deviation. 
The prior width is taken as standard deviation divided by the mean to convert from a raw number to a multiplicative factor. 
This width is calculated for neutrino data (0.04) and anti-neutrino data (0.05) separately and the slightly more conservative width from the anti-neutrinos is used. 
This prior is applied to the lower energy cross section parameter, 100 - 350 GeV, which overlaps with the accelerator data. 
By constraining the fit on this scaling factor, we can better examine the previously unmeasured space with the 350 GeV - 5 TeV scaling factor. 
The priors in the fit are summarized in the following table (all are Gaussian and centered at 0): 
\begin{center}
    \begin{tabular}{ | l | l | l |}
    \hline
    Parameter  & Prior width \\ \hline
    Barr H$^{\pm}$  & 0.15  \\ \hline
    Barr W$^{\pm}$  & 0.40  \\ \hline
    Barr Y$^{\pm}$  & 0.30  \\ \hline
    Barr Z$^{\pm}$  & 0.12  \\ \hline
    Cosmic ray model interpolation &  1.   \\ \hline
    100 - 350 GeV cross section scale  & 0.05   \\ \hline
    \end{tabular}
\end{center}

\begin{figure} \centering
    \begin{subfigure}[b]{0.32\textwidth}
        \includegraphics[width=\textwidth]{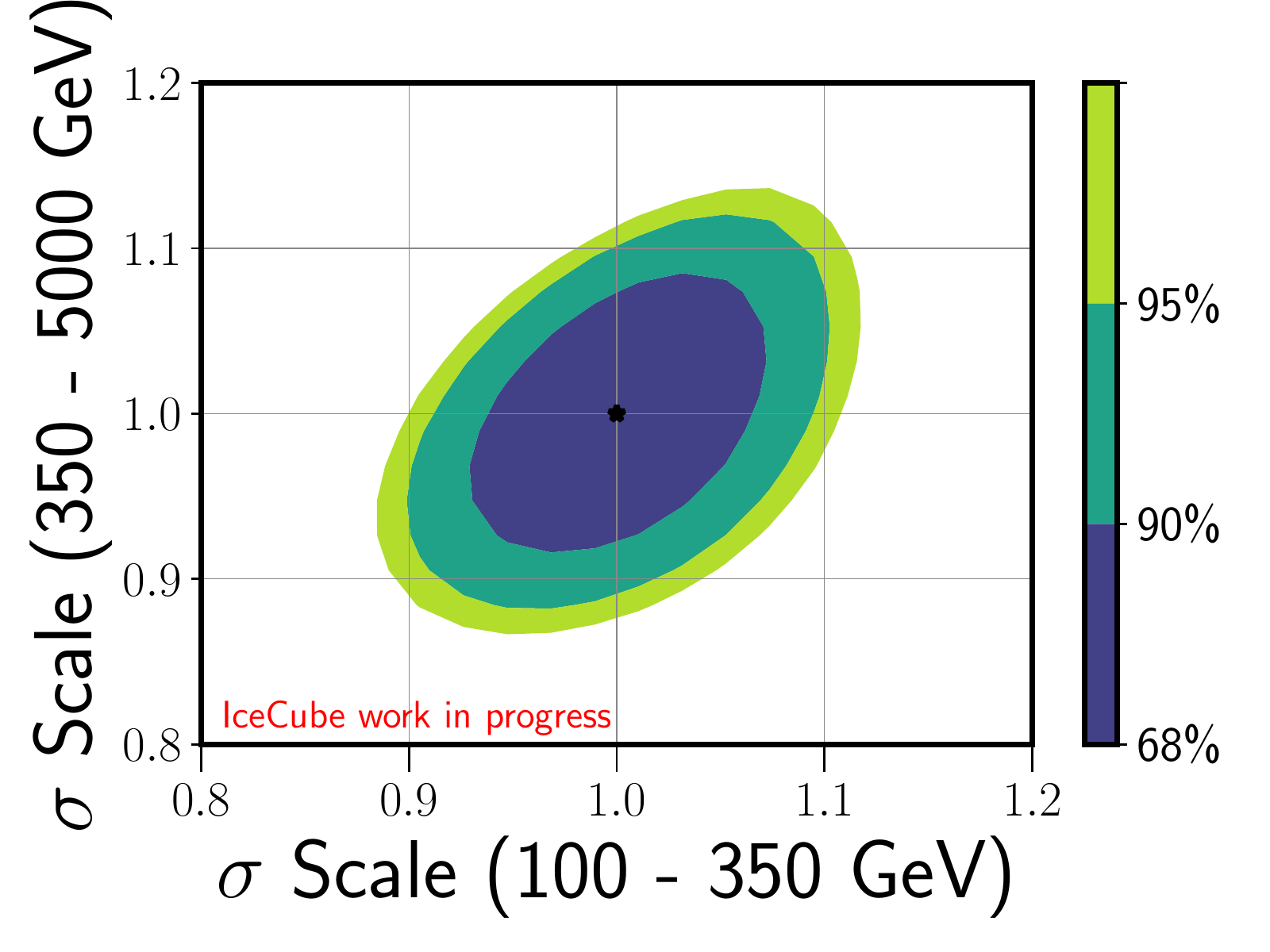}
    \end{subfigure}
    \begin{subfigure}[b]{0.32\textwidth}
        \includegraphics[width=\textwidth]{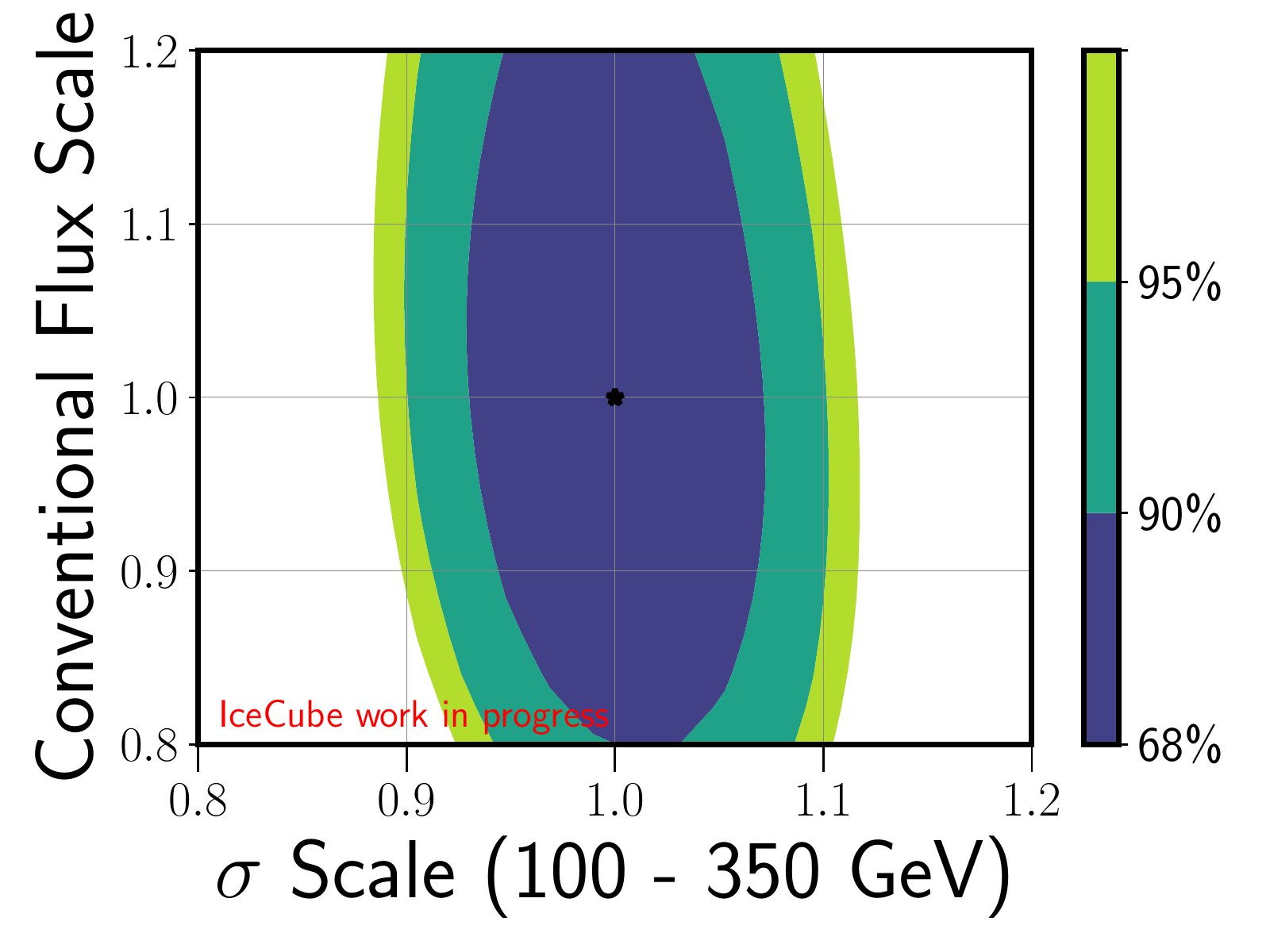}
    \end{subfigure}
    \begin{subfigure}[b]{0.32\textwidth}
        \includegraphics[width=\textwidth]{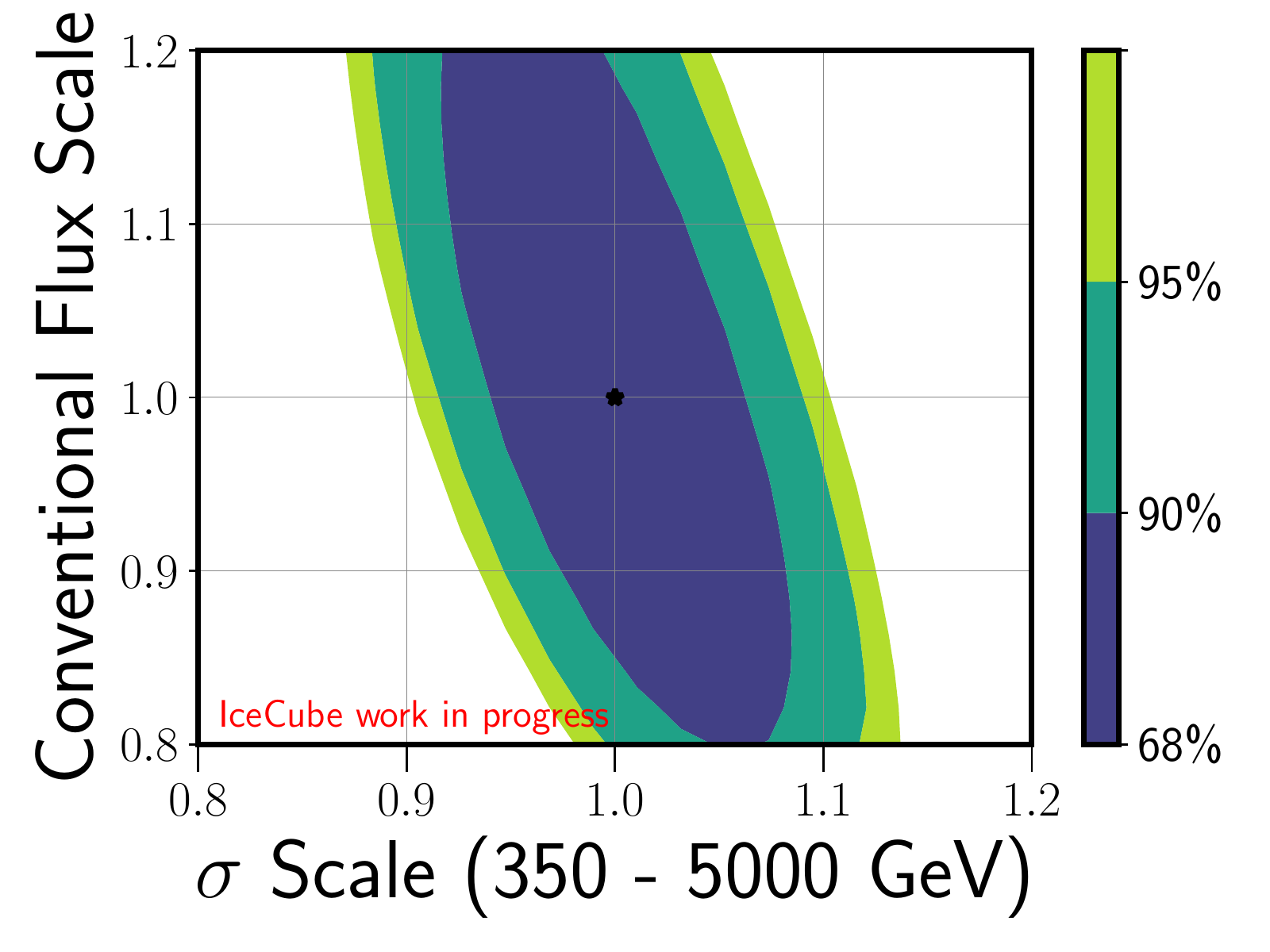}
    \end{subfigure}
\caption{Scan of profile likelihood for our two signal parameters, cross section scaling for 100 - 350 GeV and for 350 - 5000 GeV, and correlated normalization of conventional atmospheric flux. 
Note that although nuisance parameters describing variation in detector modelling are not included here, 
they are expected to have only a small effect on the contours.}
\label{fig:sensitivities}
\end{figure}

Projected sensitivities using baseline Monte Carlo and including flux-related systematic parameters look very promising, shown in Fig.~\ref{fig:sensitivities}. 
Though nuisance parameters describing detector modelling effects are not yet included, a study was run with the previous reconstruction. 
The detector nuisance parameters were largely uncorrelated with all the other parameters in the fit. 
Since we have no reason to believe that an improvement in resolution will increase the effects of these nuisance parameters, we expect the change in the sensitivities to be small. 

\section{Looking Ahead}\label{sec:future} 

To complete this analysis, there are a few remaining factors under investigation. 
Systematics varying our description of the detector will be included. 
Test will be run including detected charge in the histogram of observables, which is expected to strengthen the measurement due to the combined reconstruction. 
Studies will also be conducted concerning the use of scaling parameters for summed $\nu$ and $\bar{\nu}$ contributions, given the difference in $\sigma_{\nu}$ and $\sigma_{\bar{\nu}}$. 
For future implementations of the analysis, the very high statistics available could support additional cross section bins, providing more insight into the shape of the distribution. 
Also, though the effect is small, this work could act as the starting point for a feasibility study of a dedicated 1 - 5 TeV neutrino-Earth absorption measurement. 

In conclusion, IceCube is on the precipice of making a world-leading measurement of the neutrino-nucleon cross section in the previously unmeasured GeV to TeV energy range. 
There are plans in progress for a new accelerator experiment to measure the cross section for these energies \cite{Abreu:2021hol}, which will allow for the first cross-checks between accelerator-based and particleastro-based methods. 

\bibliographystyle{ICRC} 
\bibliography{references}



\clearpage
\section*{Full Authors List: IceCube Collaboration}




\scriptsize
\noindent
R. Abbasi$^{17}$,
M. Ackermann$^{59}$,
J. Adams$^{18}$,
J. A. Aguilar$^{12}$,
M. Ahlers$^{22}$,
M. Ahrens$^{50}$,
C. Alispach$^{28}$,
A. A. Alves Jr.$^{31}$,
N. M. Amin$^{42}$,
R. An$^{14}$
% ...
\\

\noindent
% ...
$^{12}$ Universit{\'e} Libre de Bruxelles, Science Faculty CP230, B-1050 Brussels, Belgium \\
% ...
$^{14}$ Department of Physics and Laboratory for Particle Physics and Cosmology, Harvard University, Cambridge, MA 02138, USA \\
% ...
$^{17}$ Department of Physics, Loyola University Chicago, Chicago, IL 60660, USA \\
$^{18}$ Dept. of Physics and Astronomy, University of Canterbury, Private Bag 4800, Christchurch, New Zealand \\
% ...
$^{22}$ Niels Bohr Institute, University of Copenhagen, DK-2100 Copenhagen, Denmark \\
% ...
$^{28}$ D{\'e}partement de physique nucl{\'e}aire et corpusculaire, Universit{\'e} de Gen{\`e}ve, CH-1211 Gen{\`e}ve, Switzerland \\
% ...
$^{31}$ Karlsruhe Institute of Technology, Institute for Astroparticle Physics, D-76021 Karlsruhe, Germany  \\
% ...
$^{42}$ Bartol Research Institute and Dept. of Physics and Astronomy, University of Delaware, Newark, DE 19716, USA \\
% ...
$^{50}$ Oskar Klein Centre and Dept. of Physics, Stockholm University, SE-10691 Stockholm, Sweden \\
% ...
$^{59}$ DESY, D-15738 Zeuthen, Germany \\
% ...

\subsection*{Acknowledgements}

\noindent
USA {\textendash} U.S. National Science Foundation-Office of Polar Programs,
U.S. National Science Foundation-Physics Division,
U.S. National Science Foundation-EPSCoR,
Wisconsin Alumni Research Foundation,
Center for High Throughput Computing (CHTC) at the University of Wisconsin{\textendash}Madison,
Open Science Grid (OSG),
Extreme Science and Engineering Discovery Environment (XSEDE),
Frontera computing project at the Texas Advanced Computing Center,
U.S. Department of Energy-National Energy Research Scientific Computing Center,
Particle astrophysics research computing center at the University of Maryland,
Institute for Cyber-Enabled Research at Michigan State University,
and Astroparticle physics computational facility at Marquette University;
Belgium {\textendash} Funds for Scientific Research (FRS-FNRS and FWO),
FWO Odysseus and Big Science programmes,
and Belgian Federal Science Policy Office (Belspo);
Germany {\textendash} Bundesministerium f{\"u}r Bildung und Forschung (BMBF),
Deutsche Forschungsgemeinschaft (DFG),
Helmholtz Alliance for Astroparticle Physics (HAP),
Initiative and Networking Fund of the Helmholtz Association,
Deutsches Elektronen Synchrotron (DESY),
and High Performance Computing cluster of the RWTH Aachen;
Sweden {\textendash} Swedish Research Council,
Swedish Polar Research Secretariat,
Swedish National Infrastructure for Computing (SNIC),
and Knut and Alice Wallenberg Foundation;
Australia {\textendash} Australian Research Council;
Canada {\textendash} Natural Sciences and Engineering Research Council of Canada,
Calcul Qu{\'e}bec, Compute Ontario, Canada Foundation for Innovation, WestGrid, and Compute Canada;
Denmark {\textendash} Villum Fonden and Carlsberg Foundation;
New Zealand {\textendash} Marsden Fund;
Japan {\textendash} Japan Society for Promotion of Science (JSPS)
and Institute for Global Prominent Research (IGPR) of Chiba University;
Korea {\textendash} National Research Foundation of Korea (NRF);
Switzerland {\textendash} Swiss National Science Foundation (SNSF);
United Kingdom {\textendash} Department of Physics, University of Oxford.

\end{document}